# Spreading dynamics of drops on a solid surface submerged in different outer fluids

Yingjie Fei[1,2], Qindan Zhang[1], Youguang Ma[2], Huai Z. Li [1,*]

[1] *University of Lorraine, CNRS, LRGP, F-54000 Nancy, France*

[2] *School of Chemical Engineering and Technology, Tianjin University, Tianjin 300350, China*

* Corresponding author: huai-zhi.li@univ-lorraine.fr (Huai Z. Li)

**Abstract**

**Hypothesis**

Surrounding fluids affect critically drop wetting dynamics in many applications involving viscous environments. Although macroscopic effects of outer fluid viscosity on contact line motion have been documented, the extent to which the outer fluid modulates internal flow pattern is still not well understood, largely due to experimental challenges. It is hypothesized that the external fluid exerts a dominant effect on the internal flow fields and energy dissipation, thereby altering dynamic contact angle evolution and overall wetting behavior. Elucidating this coupling mechanism is essential for advancing our understanding of multiphase spreading in complex fluid systems.

**Experiments**

We investigate the spreading of Newtonian and non-Newtonian shear-thinning aqueous drops in air versus in oil, using high-speed imaging and custom-built micro-PIV. Internal velocity and viscosity fields are measured to quantitatively relate internal flow evolution to contact line motion. Dynamic contact angle was measured and analyzed using composite model incorporating hysteresis and pinning. Scaling laws were derived to compare spreading dynamics under different outer fluid viscosities and substrate wettabilities.

**Findings**

In air, capillary waves trigger Laplace pressure gradients that drive rapid, outward internal flow as well as fast contact line motion. In contrast, viscous oils suppress wave formation and generate recirculating vortices, resulting in a significantly slower spreading process dominated by viscous drag. Despite power-law spreading in both cases, the governing timescales reflect fundamentally different mechanisms: inertial forces within the drop dominate in air, whereas external fluid viscosity controls the spreading dynamics in oil. A unified scaling incorporating outer-fluid viscosity and equilibrium contact angle gathers diverse data onto a master curve. These results underscore the central role played by outer-fluid induced internal flow in governing wetting dynamics.

# 1 Introduction

The dynamic wetting of drops on solid surfaces, characterized by the motion of the three-phase contact line, serves as a benchmark in colloid and interface science. This process is fundamentally governed by interfacial tension and its coupling with fluid flow: at equilibrium, the contact angle satisfies Young's law, while during spreading, it deviates due to viscous dissipation and other driving forces. The tight coupling among interfacial tension, viscous drag, and contact angle evolution is central to contact line dynamics.

Drop spreading, as a representative gas–liquid–solid (or liquid–liquid–solid) interfacial system, has become a theoretical paradigm for studying multiphase interfacial dynamics. In practical applications such as inkjet printing, spraying and coating, drop spreading governs coverage uniformity and pattern fidelity [1–3]. In microfluidic flows, drop spreading influences key processes including mixing, chemical reactions, and signal transduction [4–6]. In advanced manufacturing via 3D printing, it influences molding precision and interlayer adhesion [7,8]. Accordingly, drop spreading on solid surfaces has emerged as a central subject in interfacial science [9–11].

To elucidate the contact line dynamics during drop spreading, various theoretical models and scaling laws have been proposed to describe this process. In the viscosity-dominated slow spreading regime, the Cox–Voinov model [12–15] provides a quantitative relationship between the dynamic contact angle $\theta$ and the contact line advancing velocity $u$ via a capillary number ($Ca = u\eta/\gamma$) leading to a cubic scaling: $\theta^3 \sim Ca$. Under complete wetting conditions, Tanner's law describes the time evolution of the drop spreading radius as a power law: $r \sim t^{1/10}$ [16–18]. Prior to this terminal regime, an inertial regime often emerges, characterized by the scaling $r \sim t^{1/2}$, where inertia and surface tension display competing mechanism , and the spreading becomes independent of fluid viscosity [18–22]. These models have been extensively supported by numerical simulations and high-speed imaging, which together constitute the theoretical basis of drop spreading dynamics. However, these models are derived under two main assumptions: drops are Newtonian liquids and the surrounding fluid is either inviscid or exerts negligible effect, as in case with air.

However, many practical systems deviate from these classical assumptions. Drops composed of non-Newtonian fluids—such as emulsions, suspensions, polymer solutions, gels, and biological fluids—pose additional challenges to modeling. These fluids, commonly encountered in practical applications, often exhibit shear-dependent viscosity. In such systems, spatial and temporal variations of shear rate within a spreading drop give rise to evolving internal viscosity fields, potentially altering both the internal flow structure and the dynamics of the contact line. Researchers often incorporate non-Newtonian constitutive relations [23–27] and additional physical parameters—such as multicomponent friction [28] or averaged viscosity [29,30]—into Newtonian frameworks to capture the effect of complex rheology. Nevertheless, most existing studies focus on macroscopic dynamics [27,31–34], yet the internal flow fields within a spreading drop remain largely unexplored. In particular, no measurement of the spatiotemporal viscosity distribution during spreading has been reported to our best knowledge, despite its relevance to shear thinning effects and localized viscosity-controlled contact line motion.

An additional layer of complexity arises from variations in the surrounding fluid. While early studies predominantly focused on drop spreading in air, many practical applications—such as immersion coating and bioprinting—involve drop spreading in liquid environments [35]. Unlike gases, external liquids possess finite or even high viscosities, introducing substantial viscous resistance that alters contact line motion and thereby modifies the overall spreading dynamics. Although some models have incorporated the viscosity ratio between the two liquid phases and have been partially validated experimentally [15,37,38], systematic experimental data are still lacking, limiting the understanding of wetting mechanisms in liquid–liquid–solid systems. Unraveling the fundamental dynamics of such processes is crucial for the precise fabrication of soft and biological materials. This is particularly relevant to fields such as 3D printing, bioprinting, smart coatings, biomimetic lubrication, oil-water separation, and biomedical surface design—all of which critically depend on coupled dynamics at the moving contact line between internal and external fluids.

Motivated by the challenges outlined above, this study aims to establish how internal flow fields in aqueous drops relate to contact line motion, thereby elucidating the contrasting spreading mechanisms in air and oil environments. We conduct a comparative investigation of water and polymer solution drops spreading in air, mineral oil and silicone oil, using high-speed imaging combined with a custom-built micro-scale high-speed particle image velocimetry (micro-PIV) system. Through direct visualization and quantitative analysis of internal flow patterns, viscosity distributions and spreading dynamics, we reveal the coupled influence of internal and external fluids affect the evolution of the triple contact line. Based on these insights, we propose a scaling law that captures spreading behavior across diverse external fluid and fluid–substrate combinations. This framework provides a physically grounded and generalizable basis for describing complex drop spreading dynamics in the presence of outer fluids.

## 2 Materials and methods

### 2.1 Materials

Polyacrylamide powder (AN 905 SH, Lot V2581) was supplied by SNF Floerger (France). Light white mineral oil (Cat. No. 33,077-9, Lot 03412DH-397) was supplied by Sigma-Aldrich. Silicone oil (Rhodorsil 47V50) was supplied by Elkem Silicones (France). All chemicals were used as received without further purification. Deionized water (resistivity ~5.0 MΩ·cm) was prepared using an in-house purification system (Aquadem, France). Silver-coated hollow glass spheres (S-HGS-10, Dantec Dynamics, Denmark), with an average diameter of 10 μm and a density of 1400 kg/m³, were used as seeding particles.

Two types of solid substrates were used: polymethyl methacrylate (PMMA) plates

(Mitsubishi Chemical, Japan), and stainless steel plates fabricated and mechanically polished in-house. Surface morphology was characterized by scanning electron microscopy (SEM) and atomic force microscopy (AFM) (see Supplementary Information S1). The surface roughness of the stainless steel plate and PMMA plate used in our experiments was approximately 0.06 μm and 1.3 nm, respectively.

## 2.2 Methods

The schematic of the experimental setup is shown in Fig.1. The spreading dynamics investigated in this study involved both gas-liquid-solid and liquid-liquid-solid systems. In the latter system, the solid plate was submerged in a tank filled with an immiscible oil. A nozzle was fixed above the solid plate and connected to a high-precision syringe pump (PHD 2000, Harvard Apparatus, USA) to generate drops. To avoid any additional stresses associated with pendant-drop spreading on low energy surfaces, injection was stopped once the drop made contact with the solid plate. Under these conditions, as noted by Waghmare *et al.*[38], the force exerted on the drop is strictly axial in the vertical direction and does not introduce additional radial stresses at the contact line in the horizontal direction. Backlighting was applied to record the spreading process, using a high-speed camera (Phantom V711, Vision Research, USA) equipped with a macro lens (EF 100mm f/2.8, Canon, Japan) at frame rates ranging from 200 to 50,000 fps. The image processing error was approximately ±2 pixels (±15 μm). For extremely fast spreading in air, a frame rate of 50,000 fps ensured sufficient temporal resolution. The uncertainty in the spreading diameter was within ±1%, based on repeated measurements of more than three independent spreading events (Fig. S3).

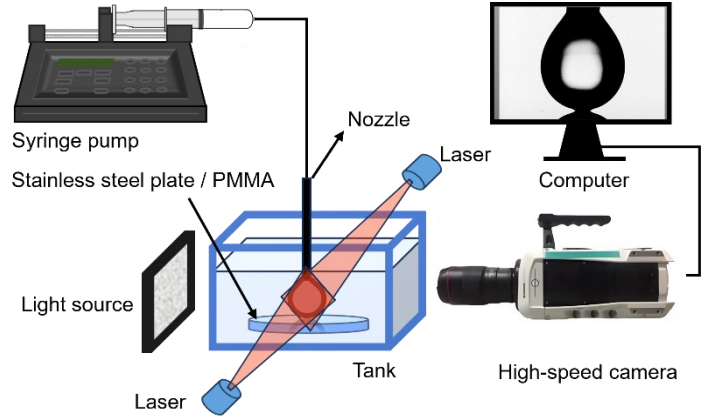

Fig. 1 Scheme of the experimental setup. The tank has a square cross section (50 mm × 50 mm) and is made of transparent PMMA. Two types of solid plates are used: stainless steel and PMMA.

We developed a high-speed micro-PIV device capable of capturing up to 5000 velocity fields per second. Two 1 mW lasers (LaserMax Inc., USA) were placed on opposite sides of the tank, with their optical paths aligned to uniformly illuminate seeding particles at the drop's equatorial plane. This setup was used to investigate the rapid spreading dynamics of polymer drops in different outer fluids. A frame rate of 5000 fps was used for spreading in air, whereas 200 fps was used in silicone oil. For the micro-Particle Image Velocimetry (micro-PIV) measurements, in addition to systematic errors, the dominant sources of uncertainty are related to the spatio-temporary resolution of seeding particles. After careful calibration and optimization, the overall error in the velocity measurements is estimated to be within approximately 1%. This small level of uncertainty is not expected to significantly affect neither the viscosity calculation within a drop nor the main conclusions of the study.

In this study, the Reynolds number of the seeding particles was sufficiently low to ensure that their motion lies within the Stokes flow regime. The Stokes relaxation time of particles ($\tau_p$) is given by $\tau_p = \rho_p d_p^2 / 18\mu$, where $\rho_p$, $d_p$, and $\mu$ are the particle density, diameter, and fluid viscosity, respectively. In PAAm solution, $\tau_p$ was calculated to be approximately 7.78 ns. Given that this timescale is orders of magnitude smaller than the characteristic flow timescale during drop spreading, the particles reliably follow the fluid motion. By carefully controlling key parameters—particle concentration, image quality, temporal interval, and spatial resolution—stable and accurate velocity vector fields were obtained.

## 2.3 Physical Properties of the Fluids

This work investigates the spreading behavior of drops on a solid plate, in which the inner fluid is an aqueous solution and the outer fluid is either air, mineral oil, or silicone oil. The inner fluids used include water and 2 wt% PAAm solution. The viscosity, density, and refractive index of all fluids used in the experiments are shown in Table 1. The interfacial tensions between each aqueous phase and the surrounding fluid are summarized in Table 2.

Table 1 Physical properties of fluids used in the experiment (20°C).

| Liquids | $\rho$ (kg·m$^{-3}$) | $\eta_0$ (Pa·s) | Refractive index |
|---|---|---|---|
| Mineral oil | 838 | 0.032 | 1.46-1.48 |
| Silicone oil | 955 | 0.056 | 1.40-1.41 |
| Water | 1034 | 0.001 | 1.34 |
| PAAm solution | 1136 | 39.440 | 1.37-1.38 |

Table 2 Interfacial tension $\gamma$ at 20°C (mN·m$^{-1}$).

| | Air | Mineral oil | Silicone oil |
|---|---|---|---|
| Water | 71.59 | 55.02 | 41.90 |
| 2 wt% PAAm | 66.20 | 54.79 | 45.26 |

The PAAm solutions are non-Newtonian and viscoelastic, exhibiting shear-thinning behavior over a wide range of shear rates. Their rheological behavior can be quantitatively described by a power-law model $\eta = K\dot{\gamma}^{n-1}$, where $K$ is the flow consistency (Pa.s$^n$), n is the flow behavior index, and $\dot{\gamma}$ is the shear rate (s$^{-1}$). For the 2 wt% PAAm solution, the fitted parameters are given by the power-law relation (Fig. S4):

$$\eta = 8.92\dot{\gamma}^{-0.63} \tag{1}$$

Table 3 lists the equilibrium contact angles $\theta_e$ (°) of water and polymer drops on stainless steel and PMMA plates. For each type of solid substrate, equilibrium contact angles were measured three outer-fluid: air, silicone oil, and mineral oil. To better match the actual spreading configuration, the equilibrium contact angles were measured by displacing the oil phase with a water-based drop. For each condition, three measurements were performed, and the average value was reported.

Table 3 Equilibrium contact angle $\theta_e$

|  | Mineral oil | Silicone oil | 2%PAAm | Water |
|---|---|---|---|---|
| Stainless steel plate / air | 17.2 | 6.9 | 54.04 | 65.57 |
| Stainless steel plate / mineral oil | - | - | 51.78 | 47.03 |
| Stainless steel plate / silicone oil | - | - | 48.10 | 47.68 |
| PMMA plate / air | 9.7 | 6.8 | 83.84 | 80.08 |
| PMMA plate / mineral oil | - | - | 119.58 | 118.32 |
| PMMA plate / silicone oil | - | - | 119.81 | 113.62 |

## 3 Results and Discussion

### 3.1 Morphological evolution

In our experiments, when the drop approached the interface in air, no measurable delay in spreading was observed, even at an imaging rate of 50,000 fps. In contrast, in an oil phase, the drop typically rested for a certain duration before initiating contact and spreading. Although the drainage of the intervening air or liquid film is also an important research topic[39,40], the present work focuses mainly on the drop spreading dynamics. To ensure that the subsequent analysis is based on the actual spreading behavior and is independent of the potentially complex film drainage process occurring prior to contact, $t = 0$ is immediately defined for the frame where the spreading radius begins to change sharply. To ensure the reproducibility of the observed phenomena, multiple experiments were repeated under identical conditions, and the results are provided in Supporting Material S2. These repeated experiments demonstrate that the variation in the onset time of spreading is statistically consistent, confirming the robustness of our definition of $t = 0$ and the reliability of the subsequent analysis.

Fig. 2 illustrates the spreading behavior of PAAm drops on a stainless-steel plate in different outer fluids, highlighting the substantial influence of the surrounding fluid. In air (Fig. 2a), initial contact between the drop and the solid substrate induces a sudden interfacial tension change that excite capillary waves[18,19,33,41,42]. These waves propagate upward along the drop surface from the contact point, generating pronounced wave crests that separate disturbed and undisturbed interfacial regions. The interface only regains a smooth and continuous profile once the waves reach the nozzle region at $t = 5$ ms. By contrast, when the substrate is immersed in an oil (Fig. 2b), viscous effects suppress interfacial disturbances, effectively inhibiting capillary wave formation and propagation. Consequently, the interface maintains an undisturbed profile throughout the whole spreading. In this case, interfacial evolution is primarily governed by changes in the liquid–liquid–solid contact angle.

The spreading width *d* is used to illustrate how solid substrate characteristics and both inner and outer fluid properties affect spreading dynamics (Fig. 2d, e). The experimental results indicate that the inner fluid, outer fluid, and substrate do not influence spreading independently; rather, their effects are strongly coupled.

In air, the influence of these factors is minimal during the early stage ($t < 1$ ms), as the spreading radii remain nearly identical. As spreading proceeds, the inner fluid viscosity and substrate wettability become more important. Higher-viscosity drops spread more slowly, and for a given fluid, spreading is slightly faster on more wettable substrates. by contrast, in oil, no early-stage universality is observed. The initial spreading rates vary significantly across the eight cases, primarily as a result of differences in outer fluid viscosity and equilibrium contact angle. Slower spreading occurs when the outer fluid is more viscous or when the equilibrium contact angle is larger.

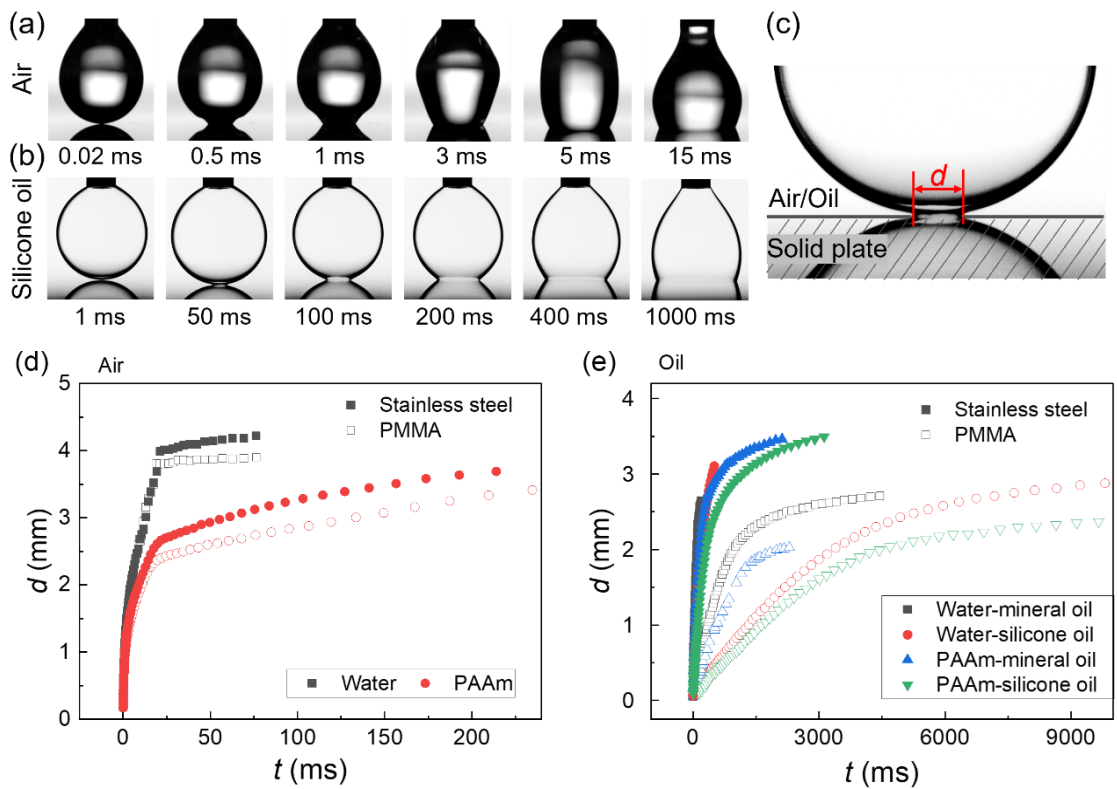

Fig. 2 PAAm Drops contact and spreading on a stainless steel plate in outer fluid. The outer fluid in a is air, in b is silicone oil. Once the drop contacts the substrate, the feeding is stopped. (c) Schematic diagram of the intersection of a drop and a solid plate. Temporal evolution of the spreading width *d*. (d) Outer fluid: air. (e) Outer fluid: oil.

### 3.2 Flow fields and viscosity distribution within a drop

The differences in spreading behavior between air and oil extend beyond the drop contour to the internal flow field. The internal flow field within the spreading drop provides a more direct and accessible reflection of the interfacial dynamics, as the flow in the surrounding fluid is induced by the spreading process[43]. To date, spatially resolved shear distribution within spreading non-Newtonian drops have not been directly quantified by experimental yet. In this study, the high-speed micro-PIV is employed to capture detailed internal flow fields (Fig. 3), enabling the first experimental estimation of viscosity distributions during spreading (Fig. 4Fig. S). The basics of the method are provided in the Supplementary Material S4.

*Velocity field*

During the drop spreading in air, the presence of capillary waves, accompanied by changes in interfacial curvature, strongly affects the internal flow field. In Fig. 3a, arrows mark the locations of the wave crests. Ahead of the crest, the interface shape remains largely unchanged, with negligible velocity vectors, indicating that the local fluid is nearly stationary and minimally involved in the bulk motion. In contrast, below the crest, the interfacial curvature reverses from inward- to outward-facing, generating a Laplace pressure gradient that drives rapid outward fluid ($t = 0.8 - 1.2$ ms). Near contact line, the maximum velocity increases from 203.9 mm/s at 0.8 ms to a peak of 252.5 mm/s at 1.0 ms, then gradually decrease to 216.0 mm/s at 1.2 ms. As the capillary wave reaches the vicinity of the nozzle ($t = 5.0$ ms), flow develops throughout the entire drop, with a maximum velocity of 76.05 mm/s.

In silicone oil, the internal flow behavior contrasts sharply with that in air. In the early stage ($t = 0 - 40$ ms), velocity fields are concentrated near the central axis and gradually form a cone-shaped downward flow. Substrate confinement, high shear, and the viscous drag of the oil suppress lateral spreading, inducing upward recirculation along the sides. This recirculation evolves into a stable, symmetric vortex, which gradually shrinks as spreading proceeds. In addition to the distinct flow pattern, the overall flow velocity is much lower than in air. The maximum velocity throughout the process, 3.25 mm/s, occurs at $t = 80$ ms in the central downward flow region, approximately 78 times smaller than the peak value observed in air.

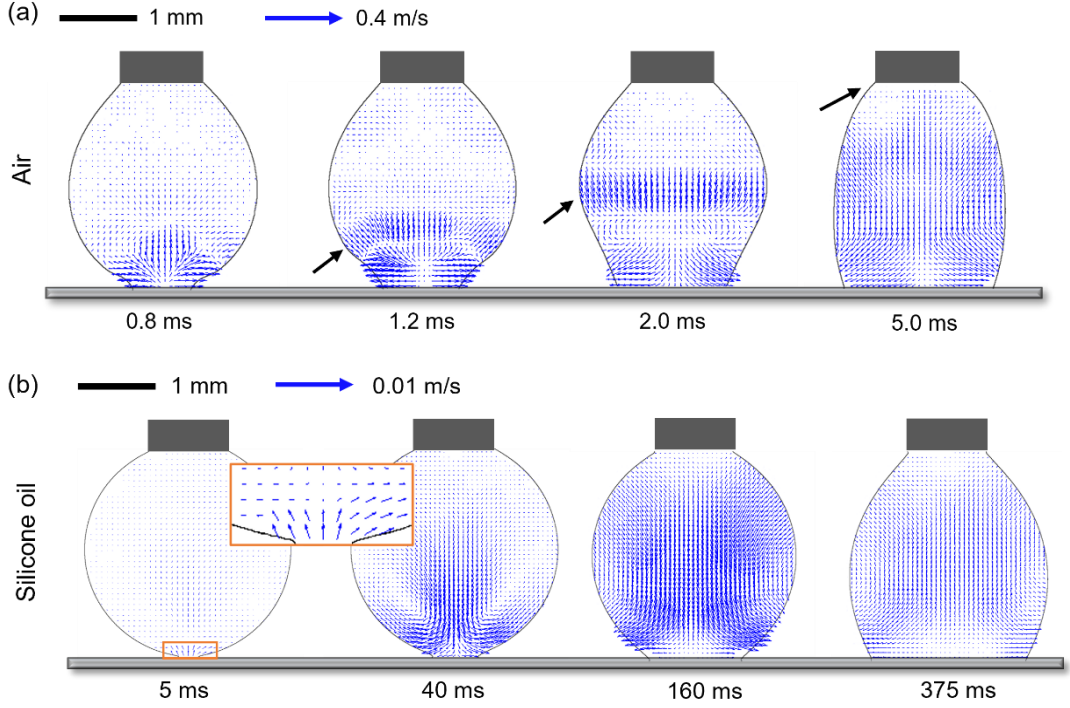

Fig. 3 The velocity field inside a spreading polymer drop in (a) is air, $D_{max} = 2.44$ mm, the spatial resolution of the flow fields is approximately 7.1 μm; The outer fluid in (b) is silicone oil. $D_{max} = 2.61$ mm. The spatial resolution of the flow fields is approximately 3.7 μm. The arrow indicates the location of the capillary wave crest. The black region represents the nozzle, and the gray region represents the solid plate.

*Viscosity distribution*

Based on the high-resolution velocity fields, we estimated the viscosity distribution within the non-Newtonian drops. Fig. 4 presents the viscosity distributions of the drop spreading in air and in oil, respectively. Overall, the viscosity field remains symmetric throughout the process. Similar to the velocity field, the viscosity distribution exhibits significant differences depending on the outer fluid.

Fig. 4a shows the case in air. Two distinct low-viscosity regions are observed: one above and one below the crest, each corresponding to different flow mechanisms. The region below the crest is driven by concentrated shear stress near the advancing contact line. At $t = 1$ ms, the shear rate near the contact line reaches approximately 549 s$^{-1}$, the highest value observed during the entire process that corresponds to a local viscosity ~ 0.15 mPa·s. As the wave propagates, this low-viscosity region remains concentrated near the contact line and gradually extends upward. The maximum shear rate decreases from ~ 210 s$^{-1}$ ($t$ = 2 ms) to ~110 s$^{-1}$ ($t$ = 5 ms), while the corresponding viscosity increases from 0.27 to 0.41 mPa·s. The second region above the crest originates from compensation flow (see Supplementary Materials S5 for details). It first appears at $t$ = 1 - 2 ms as a weak, irregular arc-shaped flow, with a maximum shear rate increasing from ~ 15 s$^{-1}$ at $t$ = 1 ms to ~50 s$^{-1}$ at $t$ = 2 ms. This region then evolves into a more stable and continuous arc-shaped trajectory ($t$ = 2.4 - 3.6 ms), where the shear rate peaks at ~170 s$^{-1}$. It eventually dissipates as the capillary wave fades ($t$ = 5 ms)

Fig. 4b shows the case in oil. Considering that the power-law model deviates significantly from the actual viscosity when the shear rate is below 0.3 s$^{-1}$ (corresponding to a viscosity of 15.41 mPa·s), the viscosity range in Fig. 4b**Erreur ! Source du renvoi introuvable.** is limited to 0 - 15 mPa·s for clarity. A symmetric, fan-shaped high-shear region initially forms around the downward conical flow and the upward recirculation zone, and gradually expands over time. The maximum shear rate increases from 3.58 s$^{-1}$ at 15 ms to a peak of 5.13 s$^{-1}$ at 25 ms, and then gradually decreases to 2.34 s$^{-1}$ by 230 ms. The evolution of the flow field reveals clearly the change in the dominant forces governing the spreading dynamics (see Supplementary Material S6).

The distinct flow patterns observed in air and oil suggest that different mechanisms may be involved in the contact line dynamics. In air, the presence of capillary waves and localized shear near the contact line could contribute to enhanced interfacial motion. In oil, the flow field is dominated by viscous recirculation, indicating that viscous dissipation and substrate wettability would play a more prominent role. These differences might lead to variations in contact line behavior, as discussed in the next section.

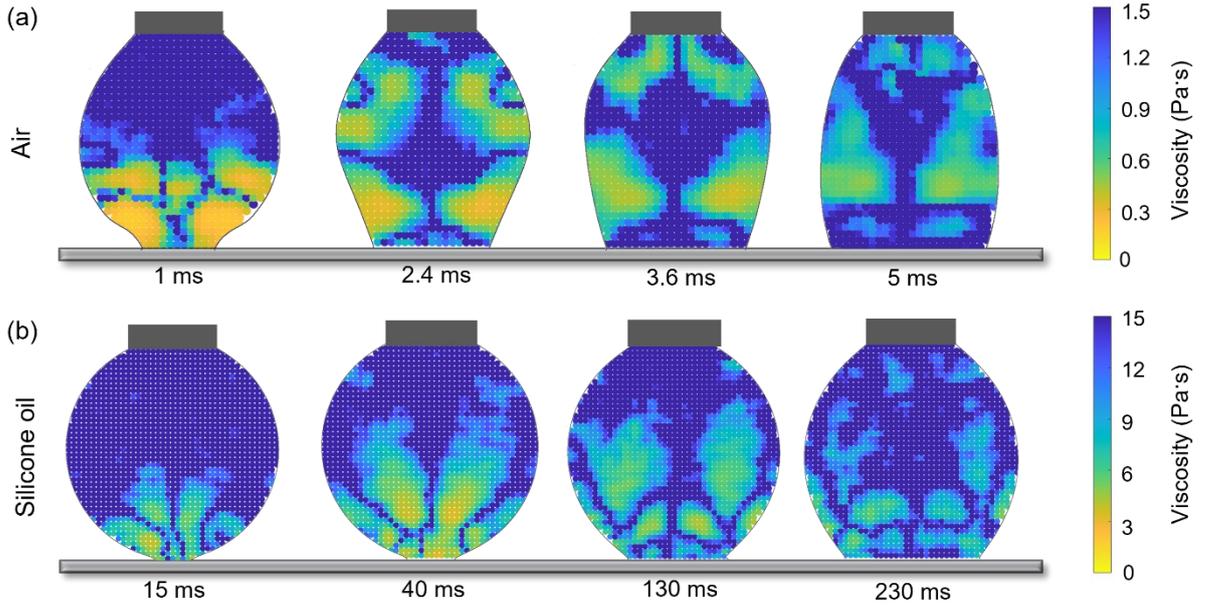

Fig. 4 Estimated viscosity $\eta_e$ distribution inside a spreading polymer drop on a stainless steel plate in (a) is air, and in (b) is silicone oil.

## 3.3 Dynamics of contact line

### 3.3.1 Contact angle–velocity relationship and model analysis

The evolution of the dynamic contact angle reflects the competition between capillary driving forces and multiple dissipation mechanisms, including viscous drag and contact line friction[9,44,45]. Moreover, it could serve as a sensitive indicator of interfacial interactions and surface heterogeneity. Chemical or geometrical inhomogeneities of the substrate can induce local pinning and depinning[46–48]. These subtle variations are directly revealed in the relationship between the dynamic contact angle $\theta_D(t)$ and the contact line velocity $U(t)$. In studying the contact line dynamics of non-Newtonian fluids, the combined effects of shear-dependent viscosity and viscoelasticity complicate the definition of $Ca$ and increase the uncertainty of model predictions. Therefore, a Newtonian fluid was used as a reference to enable direct comparison between theoretical predictions and experimental data. Fig. 5 displays the evolutions of $\theta_D$ as a function $U$ and time $t$ in air and oil, respectively.

The dynamic contact angle models generally fall into two classical approaches: the hydrodynamic (HD) model, which attributes the energy dissipation primarily to viscous drag, and the molecular kinetic theory (MKT) model, which considers the dissipation occurring at or near the contact line. As a representative of the HD model, Cox[15] proposed a classical model based on lubrication theory and the Stokes equation with the resulting expression as follows:

$$g(\theta,\varepsilon) - g(\theta_m,\varepsilon) = Ca\ \ln(x/l) \tag{2}$$

$$g(\theta,\varepsilon) = \int_0^\theta d\beta / f(\beta,\varepsilon)$$

$$f(\beta,\varepsilon) = \frac{2\sin\beta\left[\varepsilon^2(\beta^2 - \sin^2\beta) + 2\varepsilon\{\beta(\pi-\beta) + \sin^2\beta\} + \{(\pi-\beta)^2 - \sin^2\beta\}\right]}{\varepsilon(\beta^2 - \sin^2\beta)\{(\pi-\beta) + \sin\beta\cos\beta\} + \{(\pi-\beta)^2 - \sin^2\beta\}(\beta - \sin\beta\cos\beta)} \tag{3}$$

where $\varepsilon$ is the viscosity ratio between the inner and outer fluids, $\theta_m$ is the microscopic contact angle, and $Ca = \mu u/\gamma$ is the capillary number, with $\mu$ the drop viscosity, $u$ the contact line velocity, and $\gamma$ the interfacial tension. $x$ and $l$ represent appropriately chosen macroscopic and microscopic length scales, respectively, with $\ln(x/l)$ typically of the order of 10[9,45]. In the present experiments, $x/l$ is set to 1000. When the viscosity of the external fluid is negligible and the contact angle $\theta < 135°$, the function $g(\theta,0) = \int_0^\theta \frac{\beta - \sin\beta\cos\beta}{2\sin\beta} d\beta$ can be approximated as $\theta^3/9$ [12,15], leading to the classical Cox–Voinov relation:

$$\theta_D^3 - \theta_e^3 = 9Ca\ln(x/l),\ \theta_D < 3\pi/4 \tag{4}$$

Blake and Haynes proposed the molecular kinetic theory (MKT), which describes the motion of the three-phase contact line as a molecular displacement [48]. At the nanometer scale, the change in the contact angle is governed by the adsorption and desorption of liquid molecules at active sites on the solid surface[9,48]. The advancing or receding contact line is characterized by the molecular jump length and the jump frequency $\lambda$. The velocity of the contact line $U$ can be expressed as:

$$\cos\theta_D - \cos\theta_e = \frac{2nk_BT}{\gamma}\sinh^{-1}\left(\frac{U}{2K_0\lambda}\right) \tag{5}$$

Where $k_B$ is the Boltzmann constant, $T$ is the absolute temperature, and $n$ is number density of adsorption sites, which can be approximated as $n \approx 1/\lambda^2$. The contact-line friction coefficient in the MKT, $\zeta_{MKT} = k_BT/K_0\lambda$, represents the frictional force per unit length of the contact line and serves as a quantitative measure of local energy dissipation at the contact line. As shown in Fig.5a and 5c, the MKT fitting reproduces the average experimental trend well, whereas the Cox–Voinov model exhibits significant deviations from the mean behavior. This indicates that the spreading dynamics of a pendant drop are not primarily governed by viscous dissipation within the bulk fluid, but are instead dominated by molecular-scale adsorption–desorption processes at the liquid-solid interface and by contact line friction.

As shown in Fig. 5b, a sudden decrease in the contact angle was captured in the mid-to-late stage of spreading, both on the PMMA and stainless-steel substrates. This decrease is triggered by the detachment of the drop from the nozzle. On the more hydrophobic PMMA surface ($\theta_e$ = 80.08°), oscillations in the contact angle appear after the sudden decrease. This oscillatory behavior can be related to contact angle hysteresis on the substrate[49]. Since the classical MKT model assumes a homogeneous surface and lacks an explicit pinning threshold, it cannot describe the non-smooth fluctuations and rate-dependent hysteresis caused by stick–slip motion. To address this limitation, Dwivedi *et al.* [49,50] extended the classical Cox–Voinov model (Eq. 4) and the MKT model (Eq. 5) by incorporating the effects of hysteresis and pinning, and proposed a composite dynamic contact angle model:

$$\theta_D^3 = \left\{ \mathrm{acos}\left[ \cos\theta_e - \frac{\xi Ca}{\mu} - C_{\mathrm{pin}} \frac{\tanh(C \times Ca)}{\gamma} \right] \right\}^3 + 9Ca\ln(\frac{x}{l}) \qquad (6)$$

Where parameters $\xi$ and $C$ are obtained through fitting. The term $\xi Ca/\mu$ accounts for the contact line friction effect, while the term $C_{\mathrm{pin}}\tanh(C\times Ca)/\gamma$ represents the contributions of hysteresis and pinning. Here, $C_{\mathrm{pin}}$ is the pinning coefficient associated with hysteresis, defined as:

$$C_{\mathrm{pin}} = \begin{cases} \gamma(\cos\theta_e - \cos\theta_a) & U > 0 \\ \gamma(\cos\theta_r - \cos\theta_e) & U < 0 \end{cases} \qquad (7)$$

where $\theta_a$ is advanced angle and $\theta_r$ is receding angle. Fig. 5b shows a comparison between this model and our experimental data. The model successfully reproduces the sudden decrease and subsequent oscillations of the contact angle triggered by drop detachment from the nozzle. Although the magnitude of $C$ in our experiments is slightly lower than $10^4$ reported by Dwivedi *et al.*[49], its validity is confirmed in Supplementary Information S8.

When the viscosity of the surrounding fluid is no longer negligible, significant viscous stress in the boundary layer and coupled flow pattern at the liquid–liquid interface induce much more complex hydrodynamics near the contact line than that in a gas-liquid-solid system. As a result, the Cox–Voinov model is no longer applicable to liquid–liquid–solid systems. Some researchers have attempted to simplify the model by following the same formulation used for air–liquid systems (Eq. 4) [36,51], like ${\theta_D^a} - {\theta_e^a} = bCa\ln(x/l)$. However, this simplified model is highly sensitive to the contact angle range and can introduce significant errors in subsequent fitting. Therefore, inspired by Eq. 6, we define:

$$\theta_m = \mathrm{acos}\left[\cos\theta_e - \frac{\xi Ca}{\mu} - C_{\mathrm{pin}}\frac{\tanh(C \times Ca)}{\gamma}\right] \quad (8)$$

By substituting Eq. 8 into Eqs. 2 and 3, the parameters $\xi$ and $C$ were obtained through data fitting. Due to the strong nonlinearity of the model, a numerical optimization strategy combining logarithmic parameterization, multi-start search, and nonlinear least-squares fitting was employed to improve the robustness and stability of parameter estimation. To validate the model, we set $\varepsilon = 0$ (i.e., air environment) and compared the results with those from Eq. 6. As summarized in Table 4, the fitted errors are typically within 2–5% for $\xi$, and within 0.2% for $C$. This satisfactory agreement demonstrates the feasibility of the present model and optimization algorithm for describing the spreading dynamics in liquid–liquid–solid systems.

The experimental data for four liquid-liquid-solid systems were fitted using three different models, the results are summarized in Table 4. From the perspective of capillary driving force and energy dissipation, the evolution of the dynamic contact angle reflects the superposition of three competing mechanisms: contact line friction, pinning/depinning, and macroscopic viscous bending. If the two latter contributions are neglected (MKT), they are implicitly lumped into the friction term, resulting in a significant overestimation of $\xi$ by approximately 1–2 orders of magnitude. For liquid–liquid systems, the fitted $\xi$ values obtained from the present model are 25–38% lower than those from the Dwivedi *et al.* model. This difference arises because, when the external fluid viscosity is non-negligible, using the exact Cox integral (Eq. 3) enables a more accurate exclusion of the macroscopic viscous bending contribution to the dynamic contact angle. On the other hand, $C$ is primarily determined by the shape of the pinning–sliding transition at low spreading speeds and is insensitive to how viscous bending is treated. Consequently, $C$ values obtained from the two composite models are nearly identical, with differences less than 3%.

There is an intrinsic difference in the energy dissipation mechanisms governing the spreading dynamics between air–liquid and liquid–liquid systems. In air–liquid systems, the viscosity of the outer phase is negligible, and energy dissipation occurs predominantly near the contact line. As a result, contact line friction and pinning behavior directly control the $\theta_D - U$ relationship, leading to relatively low values of $\xi$ (on the order of $10^{-2}$ Pa·s) and small values of $C$ ($10^3 - 10^4$). Surface heterogeneity (e.g., stainless steel substrates) further enhances local friction. In contrast, in liquid–liquid systems, the viscosity of the outer phase makes macroscopic viscous bending the dominant dissipation mechanism. Consequently, $\xi$ increases significantly ($1 - 10$ Pa·s), and $C$ reaches $10^5 - 10^6$. The viscosity of the outer fluid plays a much stronger role in determining the overall resistance and the $\theta_D - U$ relationship. For instance, the silicone oil exhibits higher frictional resistance and pinning thresholds than the mineral oil as expected. In summary, the wetting behavior in air–liquid system more directly reflects surface characteristics, whereas in liquid–liquid systems it is strongly modulated by the viscous dissipation between these two fluid phases.

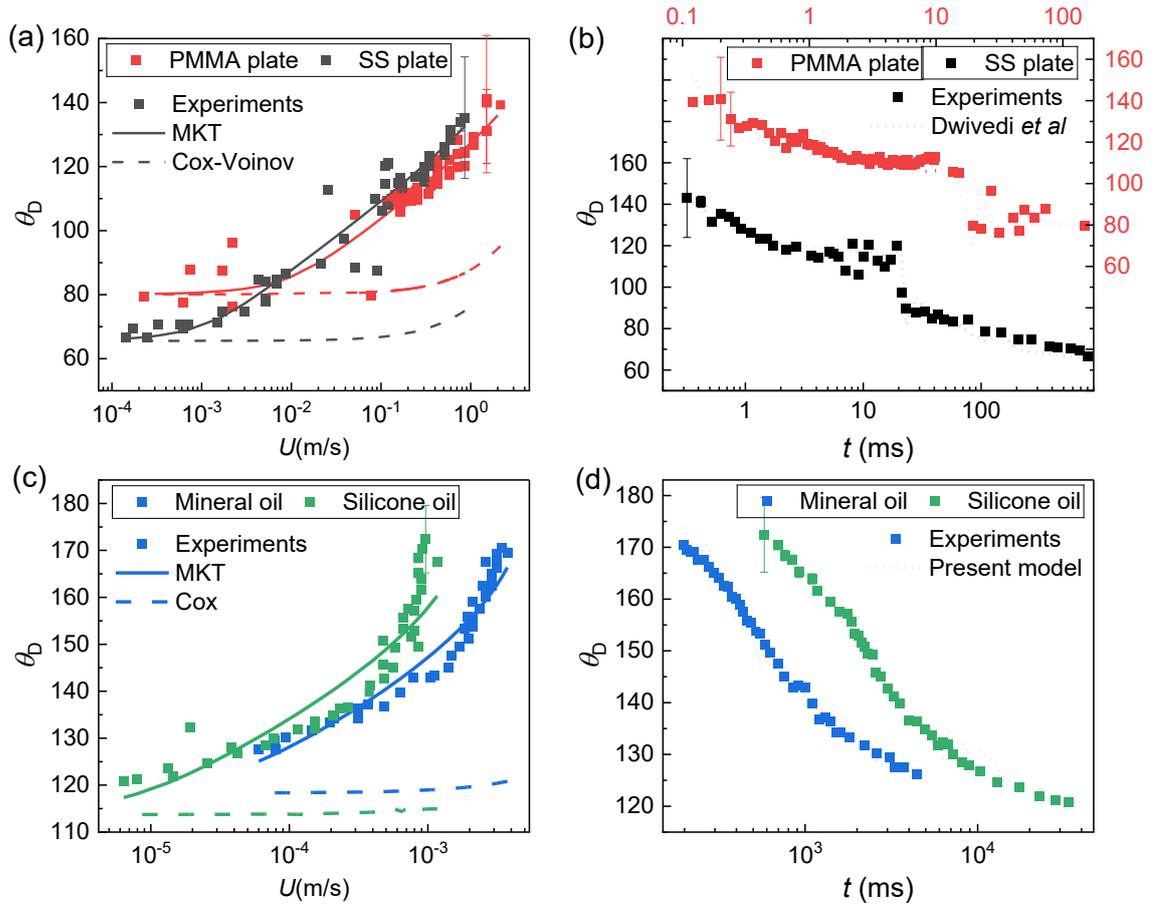

Fig. 5 Dynamic contact angle $\theta_D$ as a function of contact line velocity $U$ and time $t$. (a,b) Water drops spreading on PMMA and stainless steel (SS) plates in air. (a) $\theta_D – U$; (b) $\theta_D - t$. Experimental results are compared with the MKT, Cox–Voinov and Dwivedi *et al.* models. (c,d) Water drops spreading on PMMA plate immersed in oil (mineral oil and silicone oil). (c) $\theta_D – U$; (d) $\theta_D - t$. Experimental data are compared with the MKT, Cox models, and the present model.

Table 4 $\xi$ and $C$ obtained by applying different models to experimental data for various systems

|  |  | Contact line friction $\xi$ (Pa·s) | | | Smoothing parameter $C$ | |
|---|---|---|---|---|---|---|
|  |  | MKT | Dwivedi *et al.* (Eq. 6) | Present model (Eqs. 2,3,8) | Dwivedi *et al.* (Eq.6) | Present model (Eqs. 2,3,8) |
| Air | PMMA | 0.74 | 0.019 | 0.020 | 1335.9 | 1338.4 |
|  | SS | 5.90 | 0.052 | 0.053 | 7231.7 | 7230.4 |
| Oil (PMMA) | Mineral oil | 109.51 | 4.40 | 2.75 | $3.8 \times 10^5$ | $3.9 \times 10^5$ |
|  | Silicone oil | 402.30 | 13.74 | 10.32 | $1.7 \times 10^6$ | $1.7 \times 10^6$ |

**3.3.2 Scaling laws of spreading dynamics**

While the contact angle evolution captures key interfacial dissipation mechanisms, its predictive capability is usually restricted by the underlying assumptions of Newtonian fluid properties. Although various improvements have been proposed [31,32], they remain dependent on specific fluid properties. In contrast, the spreading diameter exhibits a power-law evolution $d \sim t^\alpha$ over a wide range of fluid systems[16–22,28,31,34,52–54]. This scaling provides a useful basis for a viscosity-independent description of wetting dynamics, serving as a bridge between contact angle behavior and neck expansion dynamics. To enable a thorough comparison between different operating conditions, time is normalized by the inertial time $t_i = (\rho D_{eq}^3/\gamma)^{0.5}$, and spreading diameter by the equivalent initial drop diameter $D_{eq}$ The dimensionless time and diameter are then defined as $\tau = t / t_i$ and $\Phi = d / D_{eq}$, respectively.

For the initial spreading in air, the dimensionless data can be fitted by a power-law with an exponent of 0.46 when $\tau < 0.03$ (Fig. 6a). This scaling is consistent with the classic inertial regime observed in drop coalescence, where the bridge radius grows as $r \sim t^{1/2}$, reflecting a balance between inertial and capillary forces: an effect well documented in previous studies[18–20,55]. However, when $\tau > 0.03$, deviations from the 1/2 exponent appear, and the exponent begins to depend on both non-Newtonian properties[27] and substrate wettability[20]. In oil, the rescaled spreading $\Phi \sim \tau^\alpha$ differs from that observed in air, and the behavior is similar to the unscaled relation $d \sim t^\alpha$ (see Fig.S7). This suggests that inertial effect of the inner fluid is inhibited by the viscous drag of the outer fluid. The observed exponent of $\sim 3/4$ implies that viscous dissipation could play a dominant role in the dynamics of liquid-liquid displacement.

For drop spreading in presence of an oil, as in the above-mentioned discussions, the viscosity of the external phase makes macroscopic viscous bending the dominant dissipation mechanism, while the equilibrium contact angle provides the primary capillary driving force. To account for these effects, we introduced a modified time scaling that incorporates these two parameters. Specifically, the influence of the external fluid is characterized by the Ohnesorge number of outer phase, defined as $Oh_o = \eta_o/(\rho_o D_{eq}\gamma)$, where $\eta_o$ and $\rho_o$ are the viscosity and density of the outer fluid, respectively. A modified dimensionless time scale is proposed: $\tau' = t / (t_i\, Oh_o \cdot \theta_e^3)$. As shown in Fig. 6c, this scaling collapses all experimental data issued from the eight different cases onto a master curve. Notably, $t_i \cdot Oh_o$ simplified to $D_{eq} \cdot \eta_o/\gamma$, which corresponds to the characteristic viscous timescale of the outer fluid $t_{v,o}$. These findings confirm that the early-stage spreading of aqueous drops in oil is governed by the viscous dissipation near the contact line. This conclusion is consistent with the molecular dynamics simulations of Primkulov *et al*.[37], and further supported by the recirculating flow patterns observed near the contact line by the micro-PIV (Fig.3b**Erreur ! Source du renvoi introuvable.**). In contrast, the initial spreading in air is mainly driven by the Laplace disjunction pressure gradient arising from interfacial curvature (Fig.3aFig. S) rather than the viscous effect [18].

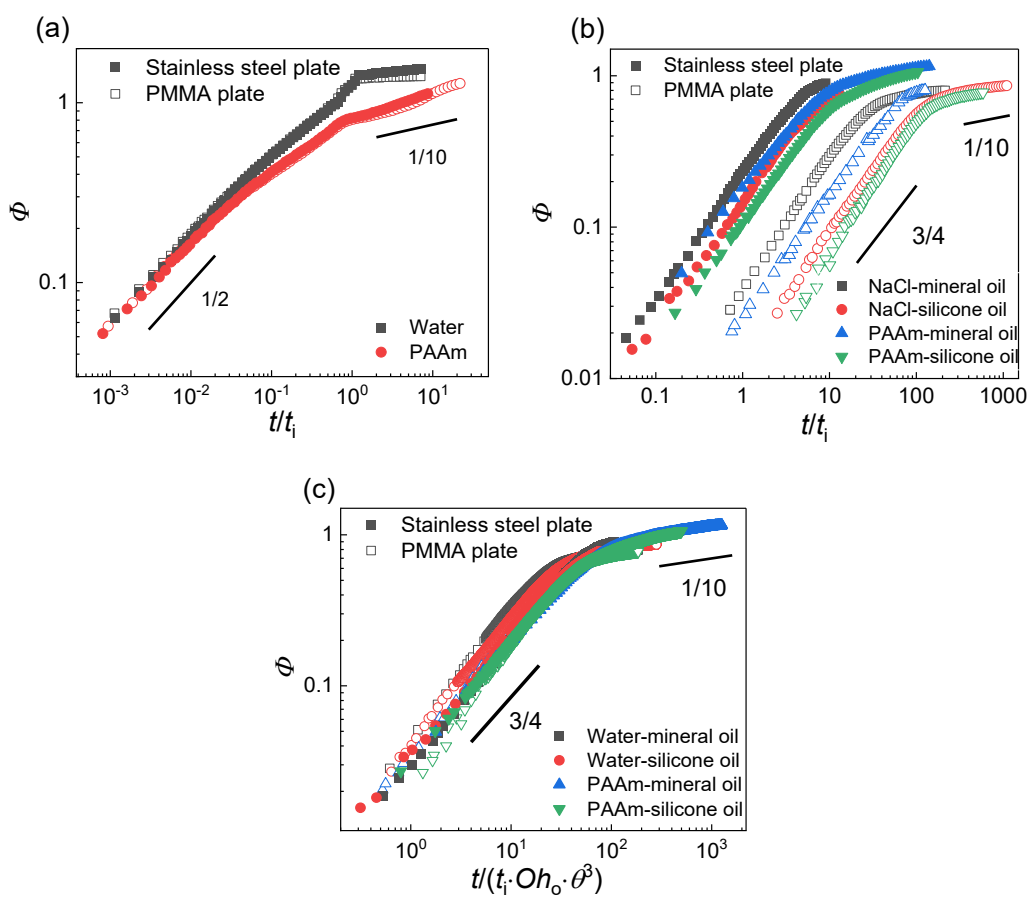

Fig. 6 Relationship between the normalized time $\tau$ ($\tau = t / t_i$) and the rescaled spreading radius $\Phi$. Drops spreading (a) in air and (b) in oil. (c) The variation of rescaled spreading radius $\Phi$ with $\tau'$ ($\tau' = t / (t_i \cdot Oh_o \cdot \theta_e^3)$) for drops spreading on solid plate in oil.

In the final stage of drop spreading, whether in air or oil, the power-law exponent of 1/10 indicates to the general tendency toward the Tanner's law. This slow spreading regime results from a balance between the capillary forces and viscous resistance near the contact line[45]. Notably, the transition from inertia- to viscosity-dominated behavior occurs at a characteristic timescale that is largely independent of fluid viscosity and equilibrium contact angle, consistent with the findings of Bird *et al*. [19]

**Conclusion**

This study reveals the fundamental role of the surrounding fluid in governing drop wetting and spreading by combining macroscopic interfacial phenomena with internal flow pattern and energy dissipation. The quantitative results by high-speed imaging and micro-PIV reveal that the outer fluid directly modulate both interfacial evolution and internal hydrodynamics. In air, the negligible viscous effect enables rapid interfacial deformation of drop. Interfacial tension imbalance at initial contact induces then capillary waves, and the resulting Laplace disjunction pressure gradient drives outward flow, with spreading mainly governed by contact line friction and pinning[49]. In oil, strong viscous drag suppresses interfacial disturbances, reorganizing the internal flow into a symmetric recirculating structure and shifting energy dissipation from local contact line effects to macroscopic viscous bending. This transition is reflected in scaling behavior: inertia-dominated $d \sim t^{1/2}$ in air [18–22] and outer fluid viscosity-dominated $d \sim t^{3/4}$ in oil.

A key innovation of this work is the first experimental quantification of spatially resolved viscosity fields during the spreading of non-Newtonian drop—an area that has remained mainly theoretical in previous literature [23–25,28–31,52]. In particular, for liquid–liquid systems, the composite dynamic contact angle model reveals that macroscopic viscous bending of the oil dominates energy dissipation, fundamentally altering the $\theta_D – U$ relationship compared to air–liquid systems [9,56]. Moreover, the modified scaling law, which incorporating the outer fluid viscous timescale and equilibrium contact angle, successfully gathers all spreading data onto a master curve, highlighting the central role of viscous coupling in governing early-stage spreading dynamics.

From the perspective of colloid and interface science, this work opens new directions for investigating and controlling dynamic wetting in complex fluid environments. It offers a generalizable framework for interpreting flow-governed spreading phenomena, with practical relevance to fields such as bioprinting, immersion coating, microfluidics, and biomedical surface design. Future works may explore the influence of substrate elasticity, surface heterogeneity, or viscoelastic outer fluids to further enrich the proposed framework and expand its application scope. In addition, coupling the experimentally observed contact angle behavior with numerical simulations such as lattice Boltzmann method (LBM) will be an interesting avenue to explore, even if the modeling of triple line dynamics remains a challenging problem for the numerical simulation.

## Acknowledgments

The fellowship provided by the China Scholarship Council to Y. Fei is greatly acknowledged.

# Spreading dynamics of drops on a solid surface submerged in different outer fluids

Yingjie Fei[1,2], Qindan Zhang[1], Youguang Ma[2], Huai Z. Li [1,*]

[1] *University of Lorraine, CNRS, LRGP, F-54000 Nancy, France*

[2] *School of Chemical Engineering and Technology, Tianjin University, Tianjin 300350, China*

* Corresponding author: huai-zhi.li@univ-lorraine.fr (Huai Z. Li)

# S1: Surface morphology of the solid substrates

The spreading dynamics investigated in this study involve both gas–liquid–solid and liquid–liquid–solid systems. Two types of solid substrates were employed: a custom-fabricated stainless steel plate and a commercial PMMA plate. Since surface roughness or defects can influence wetting behavior, particularly at small scales, scanning electron microscopy (SEM) and atomic force microscopy (AFM) were used to characterize the solid plates surface, as shown in Figure S1 and Figure S2.

The stainless steel substrate used in this study was characterized using both SEM (JSM6590-LV, JEOL, Japan) and AFM (MFP-3D, Oxford Instruments, UK) to assess its surface morphology. As shown in Figure S1a, the SEM image reveals sparsely distributed microscale features, such as shallow pits, likely resulting from mechanical polishing or machining processes. No apparent periodic structure or directional pattern are observed. Figure S1b presents a three-dimensional AFM scan over a 20 μm × 20 μm region. The surface displays a gently concave profile with height variations ranging from –0.52 μm to 1.20 μm, and no sharp protrusions or defects. The characteristic roughness length scale is significantly smaller than the length scale associated with the motion of the contact line during spreading. Therefore, although some degree of surface heterogeneity is present, it is not expected to strongly influence the overall wetting dynamics.

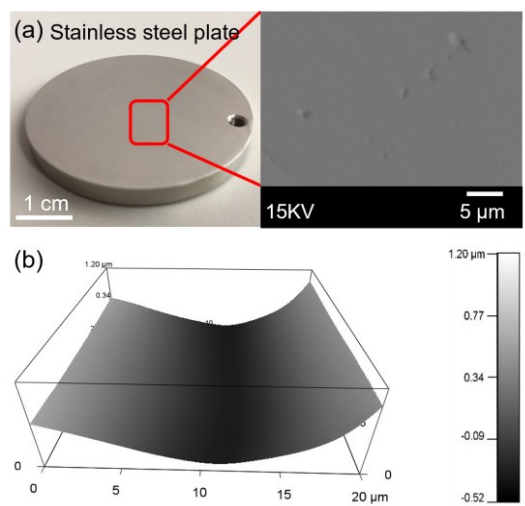

Figure S1. Surface morphology of the stainless steel substrate. (a) Scanning electron microscopy (SEM) image showing microscale surface features at 3500× magnification; (b) Atomic force microscopy (AFM) 3D topography of a 20 μm × 20 μm area.

The PMMA plate was fabricated using a high-precision mold casting method, the surface morphology was characterized using SEM (MIRA LMS, TESCAN, Czech Republic) and AFM (Dimension Icon, Bruker, Germany), as shown in Fig.S2. The SEM image reveals a relatively featureless surface at the microscale, with no distinct structures or directional textures. The AFM measurement performed over a 20 μm × 20 μm region shows surface height variations between approximately –5.5 nm and 5.6 nm. This nanoscale roughness is negligible compared to the relevant contact line length scales during spreading and is unlikely to influence the wetting behavior.

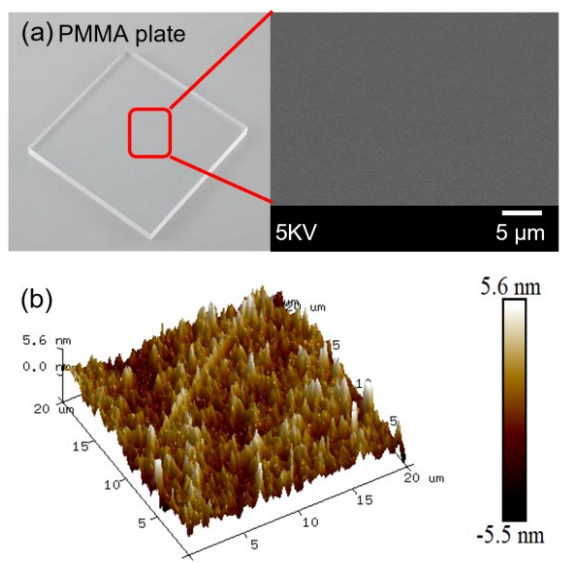

Figure S2. Surface morphology of the PMMA substrate. (a) Scanning electron microscopy

(SEM) image; (b) Atomic force microscopy (AFM) 3D topography of a 20 μm × 20 μm area.

# S2: Reproducibility of experimental results

Fig.S3 demonstrates the reproducibility of the measured spreading width $W(t)$ for two representative cases under different imaging frame rates. The left panel shows results for a water drop spreading in air on a stainless steel plate, while the right panel presents the spreading of a PAAm drop in silicone oil on a PMMA plate. In both cases, four independent measurements were performed using varying camera frame rates, ranging from 10000 to 50000 fps for the air–water system, and 1000 to 2000 fps for the oil–PAAm system. Despite differences in sampling resolution, the resulting $W(t)$ curves overlap closely across all tests, confirming the high reproducibility and robustness of the experimental methodology.

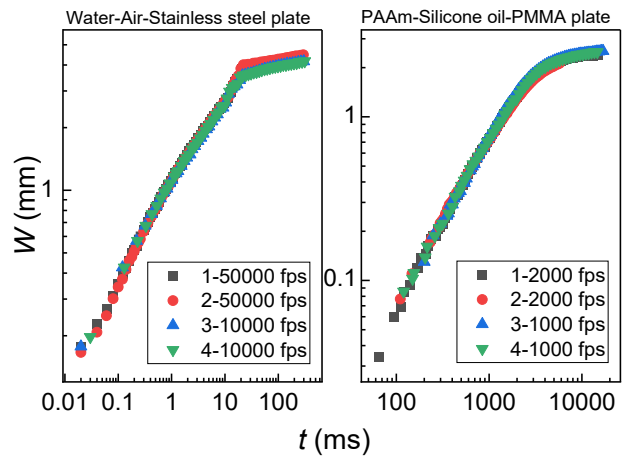

Fig.S3 Comparison of four repeated measurements of the spreading width $W(t)$ at different frame rates, showing the reproducibility of the results. Left: Water drop in air spreading on a stainless steel plate. Right: PAAm drop in silicone oil spreading on a PMMA plate.

# S3: Rheological properties of PAAm

Fig.S4 presents the shear-rate-dependent viscosity of the 2 wt% PAAm solution used in the experiments, measured at 20 °C.

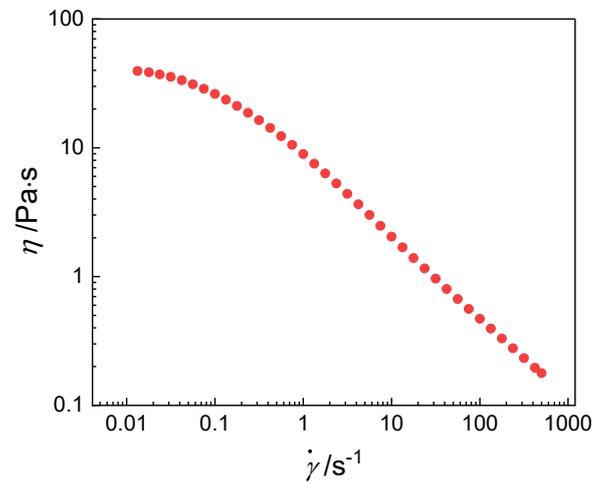

Fig. S4 Rheological properties of 2 wt% PAAm solution at 20°C.

# S4: Basics of viscosity estimation

The instantaneous viscosity distribution within the drop can be estimated from the calculated shear rate based on the experimental rheological characterization. In the cylindrical coordinates, the mean axisymmetric shear rate in the drop can be expressed as:

$$\dot{\gamma} = \frac{1}{2}\left(\frac{\partial u_r}{\partial z} + \frac{\partial u_z}{\partial r}\right) \tag{S1}$$

where $u_r$ and $u_z$ are the velocity components in the radial $r$ and axial $z$ directions respectively. The experimental flow fields obtained by the micro-PIV allow then to compute these derivatives in each direction:

$$\frac{\partial u_r}{\partial z}(m,n) \cong \frac{u_{r\,m+1,n} - u_{r\,m-1,n}}{z_{m+1,n} - z_{m-1,n}} = \frac{u_{r\,m+1,n} - u_{r\,m-1,n}}{2\Delta z} \tag{S2}$$

$$\frac{\partial u_z}{\partial r}(m,n) \cong \frac{u_{z\,m+1,n} - u_{z\,m-1,n}}{r_{m+1,n} - r_{m-1,n}} = \frac{u_{z\,m+1,n} - u_{z\,m-1,n}}{2\Delta r} \tag{S3}$$

Therefore, the resulting viscosity can be computed by the above-mentioned rheological laws.

# S5: Local velocity field evolution near the capillary wave crest

During the upward propagation of the capillary wave ($t$ = 0.8 - 5.0 ms), a distinct acceleration zone forms near the crest, moving upward with the wave. Within this region, the velocity field becomes progressively more uniform. At $t$ = 1.0 ms, the velocity at cross-section A–A (**Erreur ! Source du renvoi introuvable.**) ranges from 49.93 to 145.1 mm/s, while at $t$ = 2.0 ms, the range at cross-section B–B narrows to 70.96–127.60 mm/s.

Notably, liquid motion in opposite directions appears near this region, as shown in the magnified view in Fig.S5. The orange arrows in the figure indicate the direction of flow. By $t$ = 2.0 ms, pronounced counter-circulations develop with a layer of ~0.4 mm and local velocities ranging from 2.79 to 33.86 mm/s. Reverse flow is also observed above this zone, with a maximum upward velocity ~31.75 mm/s. This phenomenon arises from the finite-volume and incompressibility of the drop ($\nabla \cdot \vec{v} = 0$), where local acceleration must be balanced by compensating flow elsewhere to conserve the mass.

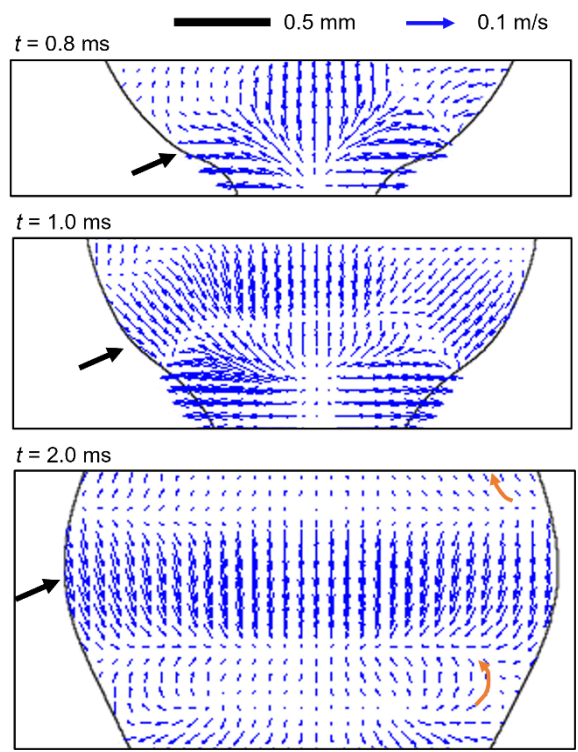

Fig. S5 Magnified view of the local velocity field inside a spreading polymer drop in air. The arrow indicates the location of the capillary wave crest.

# S6: Impact of flow structures on contact line dynamics

We sincerely thank the reviewer for this valuable suggestion. We agree that the original statement regarding the regime transition at $t = 230$ ms was not placed appropriately in the manuscript and lacked sufficient supporting evidence. Specifically, Fig. S6(a) presents the temporal evolution of the spreading diameter $d$. At $t = 230$ ms, the data clearly deviate from the early-time power-law scaling $d \sim t^{3/4}$, indicating that the dominant mechanism governing the spreading begins to change at this moment. Furthermore, Fig. S6(b) shows the shear distribution inside the drop from 130 ms to 230 ms. The shear field gradually becomes more fragmented and irregular at $t = 230$ ms, reflecting a transition in the flow structure inside the drop, consistent with the onset of viscous effects becoming increasingly important.

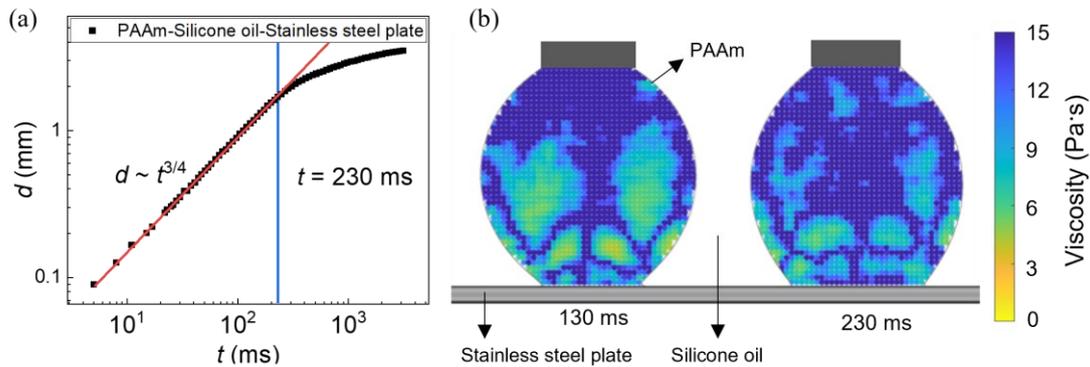

Fig. S6 (a) Temporal evolution of the bridge width $d$ between the PAAm drop and the stainless steel plate in the presence of silicone oil. The red line represents the power-law fit $d \sim t^{3/4}$ at early times, and the vertical line indicates $t = 230$ ms. (b) Estimated viscosity distribution inside a spreading polymer drop on a stainless steel plate in silicone oil at $t = 130$ ms and $t = 230$ ms.

# S7: Spreading width evolution on log–log scale: effect of fluid and substrate

Fig.S7aFig. shows the time-dependent spreading diameters of various fluid-substrate combinations in air. The log–log plot reveals that within the first 1 ms after contact, the spreading diameter of different fluids are remarkably similar. As shown in the viscosity distribution, although the viscosity near the contact line of the polymer drop is much lower than its zero-shear viscosity (0.2 mPa·s *vs*. 39 mPa·s), it is still approximately 200 times higher than that of water. This indicates that the initial spreading dynamics is largely independent of fluid property and wettability.

In contrast, the drop spreading velocity in oil shows a strong dependence on the substrate wettability and the external fluid viscosity. A clear delay in the onset of observable spreading, with differences of up to two orders of magnitude between the fastest water/mineral oil, $\theta_e$ = 47.03° and slowest PAAm/silicone oil, $\theta_e$ = 119.58° cases. Despite this variation, all eight experimental combinations follow a consistent power-law scaling behavior, with exponents ranging from 0.75 to 0.1.

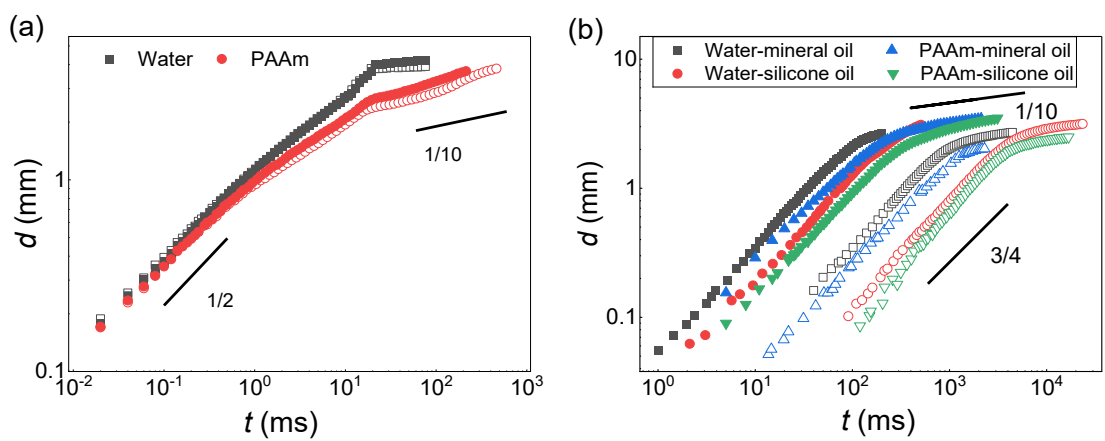

Fig.S7 Temporal evolution of the spreading radius $W$ in logarithmic scale. (a) Outer fluid: air. (b) Outer fluid: oil. The solid symbols represent the spreading on the stainless steel plate, and the hollow symbols represent the spreading on the PMMA plate.

# S8: Validation of the smoothing parameter C

To verify the validity of the slip parameter $C$, we performed an independent consistency check based on the experimentally measured contact line speed. In air, the magnitude of $C$ is typically of the order of $10^4$, as reported by Dwivedi et al. To ensure that $C$ used in our analysis corresponds to the actual speed range during the post-jump stage, we calculated the capillary number $Ca$ based on the experimental contact line velocity $U$ and fluid properties ($\mu$ = 1.0×10$^{-3}$ Pa·s, $\gamma$ = 0.072 N/m). We then determined the value of $C$ such that $\tanh(C \times Ca)$ reaches approximately 99% of its terminal value within this speed range.

$$C \approx \frac{\mathrm{atanh}(0.99)}{Ca^*} \approx \frac{2.65\gamma}{\mu U^*} \approx \frac{0.1908}{U^*} \tag{S4}$$

The corresponding characteristic contact line speeds $U^*$ for PMMA ($C_{\mathrm{PMMA}}$ = 1335.9) and stainless steel ($C_{\mathrm{SS}}$ = 7231.7) are approximately 0.14 m/s and 0.026 m/s, respectively, which are consistent with the experimentally observed speeds during the post-jump regime. Although $C$ is often used as an empirical fitting parameter in literature, this analysis confirms that the value adopted in this work effectively captures the speed range at which the jump occurs, thereby providing a physically meaningful slip parameter.